# Code in Motion: Integrating Computational Thinking with Kinematics Exploration

Mateo Dutra[1], Álvaro Suárez[2] and Arturo C. Martí[1]
[1]Instituto de Física, Facultad de Ciencias, Universidad de la República, Uruguay
[2]Consejo de Formación en Educación, Uruguay

*Although physics has become increasingly computational, with computing even being considered the third pillar of physics [1], it is still not well integrated into physics education [2]. Research suggests that integrating Computational Thinking (CT) into physics enhances conceptual understanding and strengthens students' ability to model and analyze phenomena [3]. Building on this, we designed a didactic sequence for K9 students to foster specific CT practices while reinforcing fundamental kinematics concepts. The activity revealed students' ability to apply CT skills and is well suited for use in introductory kinematics courses.*

### 1. Computational Thinking

Computation has evolved from a supporting tool into an essential component of scientific inquiry. CT is a 21st-century skill that offers a structured approach to problem-solving beyond mere programming, defined as “the thought processes involved in formulating a problem and expressing its solution(s) in such a way that a computer-human or machine-can effectively carry out” [4].

Physics education now incorporates CT into high school and undergraduate courses, as reflected in the Next Generation Science Standards (NGSS), which promote mathematical and computational thinking as core STEM practices [5]. By engaging in computational modeling, students develop both conceptual understanding and problem-solving skills.

Following Weller et al.'s framework [6], we structured a series of activities intended to progressively introduce key CT skills within the physics curriculum, such as decomposing, translating physics into code, algorithm building, debugging, applying conditional logic, generating and analyzing data, selecting effective data representations, and working in groups on computational models. Weller et al. categorize these and related practices as discipline-specific computational thinking skills relevant to physics modeling and analysis. The activity described here is part of that broader curricular effort. In order to evaluate how students engaged with these practices, we created an assessment rubric that reflects the competencies emphasized in the design of the activity and the wider sequence.

Recent research in both computational thinking and computational literacy has highlighted the particular pedagogical value of describing and interpreting code within physics instruction. For example, Gambrell and Brewe [7], based on interviews with academic and industrial physicists, report that many participants consider the ability to read and interpret code—including explaining what a program does, identifying embedded physics concepts, and justifying outputs—as important skills for students in introductory physics courses. Odden and Zwickl [8] similarly emphasize that engaging with existing code—by adapting and repurposing it—can help students construct disciplinary understanding, a perspective grounded in the concept of computational literacy. These interpretive practices are embedded in several aspects of Weller et al.'s framework—particularly in competencies such as translating physics into code, algorithm building, and applying conditional logic—and align with the goals of our activity, which integrates tasks that help students connect code behavior with physical meaning.

## 2. The activity

The didactic sequence presented here was implemented as part of the broader curricular effort to integrate CT into the physics classroom described above. We outlined our objectives and reviewed topics previously covered with the students, tailoring various activity proposals to our context. We designed and iteratively refined these proposals to best meet our students' needs. Additionally, we drafted task instructions and designed evaluation rubrics that reflected the progression of CT competencies targeted across previous and current activities.

We engaged three groups of K9 students (two groups of 21 and one of 24), all from the same school and within the same academic year, who already had a foundation in basic kinematics—including motion with constant velocity and constant acceleration, and the analysis of position–time, velocity–time, and acceleration–time graphs corresponding to such motions—as well as elementary Python programming (covering variables, data types, lists, functions, conditionals, loops, and graphs).

The activities were implemented over two 2-hour class sessions within the same week, and all stages of the sequence were carried out during regular classroom hours. Prior to this sequence, students had already engaged in other activities integrating computational thinking and physics, including tasks structurally similar to those presented here but focused on modeling motion with constant velocity. In this context of progressive integration of CT into physics instruction, the third stage of the sequence served as an opportunity to evaluate students' ability to interpret and apply CT practices in the context of modeling accelerated motion.

In the **first stage** of the activity, students worked in subgroups of four to simulate the free-fall motion using Python. Their tasks included defining functions to calculate position and velocity from the standard equations of motion, organizing time, position, and velocity data into lists, and generating graphs to visualize the results. This phase was structured as follows:

1. *Programming the Free-Fall Function*
   Students defined a Python function that accepts initial height, initial velocity, and time as inputs, and calculates the object's position and velocity using the equations:

$$x(t) = x_i + v_i t + \frac{at^2}{2} \quad (1)$$

$$v(t) = v_i + at \quad (2)$$

   This step involves translating physics into code.

2. *Testing the Function*

Students tested the function by inputting various values and printing the outputs to verify accuracy, with debugging playing a crucial role in identifying and correcting errors.

3. *Simulating a Specific Case*
   They simulated a free fall from a height of 3.8 m with zero initial velocity, generating tables for time, position, and velocity using a 0.05-second time step. This task required decomposing the problem into smaller steps, algorithm building and applying conditional logic.

4. *Graphical Representation*
   Finally, they were tasked with creating graphs position versus time and velocity versus time, based on the same initial conditions used in the previous step. Choosing how to represent the data effectively requires data representation skills, helping students connect their numerical results to visual interpretations.

In the **second stage**, the focus shifted to calculating fall time. Students derived a general formula for the fall time of an object (assuming zero initial velocity) and implemented it in Python. They applied this function to calculate the theoretical fall time for a 3.8 m drop. Then, they conducted 30 experimental trials to measure the actual fall time, computed the mean and standard deviation, and compared these results with the theoretical value to assess the validity of the free fall model used to calculate the time.

The **third stage** was an individual activity in which students analyzed a provided Python program that simulates free fall from a height of 100 m with zero initial velocity. The program calculates position using the equation

$$x(t) = x_i + \frac{at^2}{2} \tag{3}$$

with $a = -9.8 m/s^2$. It then uses a while loop to display the object's position at one-second intervals from $t = 0$ to $t = 5s$. Students were asked to provide a written explanation of the program's function and its connection to the physics of free fall (see Figure 3).

```
def free_fall(time):
    initial_position = 100
    initial_velocity = 0
    aceleration = -9.8
    position = initial_position + (initial_velocity * time) + (aceleration * (time ** 2)/2)
    return position

current_time = 0
final_time = 5
while current_time <= final_time:
    position = free_fall(current_time)
    print("At time", current_time, "s, the position is", position, "m")
    current_time += 1
```

Figure 3: Program used to assess student’s CT.

This stage was designed to assess specific CT skills—particularly those related to describing and interpreting code—that had been progressively introduced in prior activities throughout the course.

### 3. Assessing Students' Computational Thinking

To assess this subset of CT skills, we used a five-level rubric designed to capture key competencies emphasized throughout the activity. Specifically, the rubric focused on students' ability to translate physics into code, engage in algorithm building, generate data, and apply conditional logic—competencies drawn from Weller's framework and consistent with the broader instructional emphasis on integrating CT into physics. The levels of performance for each of these dimensions are described in Table 1.

Table 1: Evaluation rubric.

| Proficiency level | Description | Number of students |
|---|---|---|
| Outstanding (5) | Provides precise, comprehensive explanations that demonstrate deep mastery: clearly translates the physical model into code, detailing the iterative algorithm (algorithm building and generate data) and conditional logic used in the simulation. | 30 |
| Significant (4) | Exhibits very good understanding with clear explanations of how physics is translated into code, recognizing the iterative algorithm and conditional logic, despite minor inaccuracies. | 16 |
| Moderate (3) | Shows a solid yet incomplete understanding by identifying the conversion of physical equations into code and inferring an iterative algorithm, with only implicit mention of conditional logic. | 4 |
| Limited (2) | Provides superficial explanations that partially address CT elements, offering vague descriptions of the algorithm and data generation processes. | 9 |
| Minimal (1) | Presents explanations that are disconnected from the actual functionality, failing to recognize how the physical model is translated into code or how the underlying algorithm operates. | 7 |

At the Minimal level, students' descriptions were largely disconnected from the program's functionality and failed to address fundamental CT skills. For example, one student stated: *"What this code does is similar to a person running or moving, where the code gradually increases the velocity. As we can see, it refers to the position and starts adding the initial position and other elements. Below, it states that the position is 1, meaning it is calculated by adding the velocity, initial position, and acceleration over time to determine the current time."*

There is no *“person running or moving”* in the program, nor is the position set to 1. Another student claimed: *“We create a function that is used to calculate the fall time of an object.”* These responses indicate a lack of recognition of how CT competencies—such as translating physics into code and structuring algorithms—should be demonstrated.

At the Limited level, students provided vague or partially incorrect explanations that touched on CT elements but did not fully develop them. For instance, one student described the program as: *“The purpose of this code is to calculate the free fall experienced by an object when released from a certain height, velocity, etc. The code carries out all the necessary steps to perform this calculation. First, the variable free_fall is defined as a function of time. Then, the initial_position, initial_velocity, and acceleration are defined. Afterward, the code instructs Python on how to perform an operation to determine the next position value.”* While this response shows that the student recognized the need to compute position values, it lacks a clear explanation of the iterative algorithm and conditional logic—essential CT competencies—failing to fully articulate how the physics is translated into code.

At the Moderate level, students demonstrated a solid yet incomplete understanding of CT principles. For example, one student explained: *“This code is used to determine the position at each moment in time. Starting with the first part of the code, it begins with def free fall, followed by specifying the values for the initial position, initial velocity, and acceleration. The def free fall is used to store everything written below it, and it then explains the formula to know the position at each moment in time. Next, the code assigns different values to current time and final time, showing that the position is equal to free fall current time. Finally, it includes a print statement to display what will appear when the program is run, and that's where it ends.”* This explanation correctly identifies the conversion of a physical equation into code and infers the use of an iterative algorithm. However, the implicit treatment of conditional logic and the occasional ambiguity in terminology suggest an incomplete grasp of the full spectrum of CT competencies.

At the Significant level, students demonstrated a very good understanding of CT concepts, though with minor imprecisions. One student explained: *“The purpose of this code is to determine the position of an object at each moment in time during free fall. In other words, the program is designed to calculate and return x(t) (position as a function of time) for the time interval between 0 and 5 seconds as the object undergoes free fall. This is achieved using the return variable. In the program, a variable called free_fall is created, and all the necessary data is provided to calculate this value through a function that defines the equation used to determine the position at each moment.”* This response clearly shows how the physical model is translated into a computational function and recognizes the role of the iterative algorithm and conditional logic. Minor confusion—such as referring to a function as a variable—slightly detracts from the overall CT competency demonstrated.

At the Outstanding level, students provided excellent, detailed explanations that fully addressed CT competencies. For example, one student stated: *“In summary, this code creates a function to calculate the position of an object in free fall at a given moment in time. Then, for each time interval (one second in this case), it prints the corresponding time and position, which is calculated using the previously mentioned function. What relates this code to free fall is that the acceleration is 9.8 $m/s^2$.”* This description accurately captures how the free-fall equation is implemented in code, detailing the iterative algorithm and conditional logic while clearly linking the generated data to the physical phenomenon—demonstrating a comprehensive grasp of CT competencies.

Overall, more than 75% of the students achieved at least a Moderate proficiency, with 45% reaching the Outstanding level. These results suggest that a majority of students were able to effectively engage with the computational approach to modeling accelerated motion.

**4. Conclusion**

The results of this activity suggest that students were able to apply CT skills to analyze accelerated motion. Specifically, they successfully generated position-time and velocity-time graphs for free fall scenarios and calculated the falling time of an object using its initial position and velocity. This involved decomposing, translating physics into code, algorithm building, applying conditional logic, generating and analyzing data, debugging and working in groups on computational models.

The results obtained by the students in the individual evaluation activity highlight their ability to apply CT skills to analyze accelerated motion. With over 75% of participants achieving a sufficient level (3 or higher) and 45% reaching the highest level, the findings demonstrate their comprehension of position calculations in accelerated motion, as well as their effective use of functions and loops in Python.

While these results are encouraging, they were expected, as students' performance in Stage 3 closely reflected their success in previous tasks involving similar CT practices applied to constant velocity motion. However, the final task did not simply replicate earlier activities. Instead, it presented students with a new situation in which they were required to independently interpret unfamiliar code and connect it to a physical model. This demanded a higher level of autonomy and conceptual integration. In this sense, Stage 3 served as a meaningful assessment of whether students could transfer and apply their knowledge in a novel context and demonstrate a capacity that is often associated with deeper learning.

In Stage 3 of the activity sequence, a possible improvement for future implementations, it would be valuable to include a stopping condition in the simulation that prevents the object from continuing once it reaches the ground. This would open up the opportunity to discuss the role and limitations of mathematical models. While the current code allows the object's position to become negative, introducing such a condition could enrich the pedagogical value by prompting students to reflect on the assumptions behind the equations and how to adapt them to more realistic scenarios. In line with this adjustment, the rubric should also be updated to indicate that the highest level of performance is achieved when students recognize that the model is no longer valid once the object's position becomes negative.

Although we did not conduct a specific assessment of whether these activities improved students' understanding of kinematic concepts, existing research indicates that integrating computational thinking into physics instruction can support conceptual learning. For example, studies using modeling-based learning environments have shown that students make significant gains in kinematics when they build and simulate models of physical phenomena, especially when supported with appropriate scaffolding and activity progression [9]. Other research has highlighted that the integration of CT and physics fosters a synergistic relationship, where physics provides a mathematically grounded context for CT development, while CT practices help streamline and deepen the understanding of physics concepts [7]. Overall, synergistic learning of CT and STEM has been found to enhance students' comprehension of science content while simultaneously developing their computational skills [10]. Building on these findings, we are currently carrying out further studies to examine in greater detail how activities like the one presented here influence students' conceptual understanding in physics.

In summary, the implementation of this didactic sequence suggests that integrating computational thinking into kinematics education can enhance students' computational thinking skills and may also contribute to their understanding of theoretical concepts. Based on these findings, we suggest continuing to develop and implement similar activities, extending their application to other areas of physics to further explore the benefits of computational modeling in science education.

**References**

[1] Skuse, B. (2019). The third pillar. Physics World, 32(3), 40.

[2] Caballero, M. D., & Merner, L. (2018). Prevalence and nature of computational instruction in undergraduate physics programs across the United States. Physical Review Physics Education Research, 14(2), 020129.

[3] Orban, C. M., & Teeling-Smith, R. M. (2020). Computational thinking in introductory physics. The Physics Teacher, 58(4), 247-251.

[4] Wing, J. (2017). Computational thinking's influence on research and education for all. *Italian Journal of Educational Technology*, *25*(2), 7-14.

[5] National Research Council. (2012). A framework for K-12 science education: Practices, crosscutting concepts, and core ideas. National Academy of Sciences.

[6] Weller, D. P., Bott, T. E., Caballero, M. D., & Irving, P. W. (2022). Development and illustration of a framework for computational thinking practices in introductory physics. Physical Review Physics Education Research, 18(2), 020106

[7] Gambrell, J., & Brewe, E. (2024). Analyzing interviews on computational thinking for introductory physics students: Toward a generalized assessment. Physical Review Physics Education Research, 20(1), 010128.

[8] Odden, T. O. B., & Zwickl, B. (2025). How physics students build computational literacy by creating computational literature. Journal of the Learning Sciences, 1-43.

[9] Basu, S., Biswas, G., Sengupta, P., Dickes, A., Kinnebrew, J. S., & Clark, D. (2016). Identifying middle school students' challenges in computational thinking-based science learning. *Research and practice in technology enhanced learning*, 11, 1-35.

[10] Hutchins, N., Biswas, G., Conlin, L., Emara, M., Grover, S., & Basu, S. (2018, November). Studying synergistic learning of physics and computational thinking in a learning

by modeling environment. In *Proceedings of the 26th International Conference on Computers in Education. Philippines: Asia-Pacific Society for Computers in Education.*

**Authors**

Mateo Dutra teaches physics at the Universidad de la República (Uruguay) and physics and computer science at the high school level. He is pursuing graduate studies in PER, focusing on computational modeling and digital games for learning physics.

Álvaro Suárez teaches physics at the Instituto de Profesores Artigas (Uruguay). His research focuses on physics education, especially on students' conceptual difficulties in electromagnetism and on research-based instructional design.

Arturo Martí is a professor of physics at the Universidad de la República (Uruguay). His research spans topics from chaos and turbulence to physics education, with a special interest in the use of smartphone sensors and engaging hands-on experiments.